\documentclass[preprint,review]{elsarticle}

\usepackage{lineno,hyperref}
\usepackage{lipsum}
\modulolinenumbers[5]

\journal{Journal of \LaTeX\ Templates}

\begin{document}

\begin{frontmatter}

\title{Measurements of the neutron absorption in supermirror coatings}

\author[1]{D.D.DiJulio}\corref{mycorrespondingauthor}
\author[1]{V. Santoro}
\author[2,3]{A. Devishvili}
\author[1]{A. Khaplanov}
\author[4,5]{R. Kolevatov}
\author[6]{M. Mag\'an}
\author[1]{T.M. Miller}
\author[1]{G. Muhrer}

\address[1]{European Spallation Source ERIC, Lund Sweden}
\address[2]{Division for Materials Physics, Department of Physics and Astronomy, Uppsala University, Sweden}
\address[3]{Inst Laue Langevin, CS 20156, F-38042 Grenoble, France}
\address[4]{Department for Neutron Materials Characterization, Institutt for Energiteknikk, Instituttveien 18, Postboks 40, NO-2027 Kjeller, Norway}

\address[5]{Department of Chemistry, Aarhus University, Langelandsgade 140, 8000 Aarhus C Denmark}

\address[6]{ESS-BILBAO Parque Tecnológico Bizkaia Laida Bidea, Edificio 207 B Planta Baja, 48160 Derio, Spain}
\cortext[mycorrespondingauthor]{douglas.dijulio@ess.eu}

\begin{abstract}
In this work we report on measurements of neutron absorption in supermirror coatings. The measurements were carried out using the SuperADAM instrument at the Institut Laue-Langevin and by measuring the gamma-ray production from $m=3$ and $m=4$ neutron supermirrors when illuminated by a beam of neutrons. The results provide a valuable validation for recent computational and theoretical work that can be used as input to Monte-Carlo radiation transport calculations for the design of the shielding of neutron scattering instruments. 
\end{abstract}


\end{frontmatter}


\section{Introduction}
At a spallation neutron source, such as the European Spallation Source (ESS) \cite{Garoby_2017}, currently under construction in Lund, Sweden, the thermal and cold neutrons to be used for neutron scattering experiments will originate from the bombardment of a heavy metal target with a high-energy proton beam. The secondary neutrons created during this process are slowed down to the thermal and cold energy regime by moderators \cite{zanini2019} placed around the target location and guided to the positions of the neutron scattering instruments, which can be up to or around 150 meters from the source position, using neutron supermirror guides \cite{Mezei1976}. 
Neutron supermirrors consist of alternating layers, such as Ni and Ti, and are characterized by the critical angle, $\theta_c=m\theta_{c}^{Ni}$, where $\theta_{c}^{Ni}$ is the critical angle of pure nickel. Neutrons which are incident on the supermirror surface, with an angle less than the critical angle, will be reflected with a high probability. Thus, the higher the critical angle, the more efficient the supermirror is at reflecting neutrons. Recently, results have been reported where supermirrors have been fabricated with $m=8$ \cite{Schanzer2016}. 

A major component of the neutron scattering instruments at a neutron source is the shielding required for both radiation safety at the facility and also for improving the performance of the instruments through background reduction \cite{cherkashyna2014,Santoro:2018fid,background}. Far from the source position, and in the absence of high-energy secondary radiation from the source, the shielding requirements along the instruments are driven by the gamma-ray production along the beamline of the instrument. This arises partially from the absorption of the non-reflected thermal and cold neutrons in the supermirror coatings of the neutron guides. In the case of Ni, for example, this can lead to the production of gamma-rays with energies up to the order of 9 MeV \cite{ersez2006,jensen2014}. Thus one could expect that an accurate modelling of the gamma-ray production along the neutron beamline during the design phase would translate into a cost-savings for the facility, in addition to providing more accurate estimates of the prompt dose rates in the vicinity of the beamline shielding. 

The shielding design of neutron scattering instruments is typically carried out using Monte-Carlo codes such as PHITS \cite{niita2006}, the MCNP family of codes \cite{mcnp}, Geant4 \cite{Geant4,agostinelli2003} and FLUKA \cite{fluka}. However, of these codes, only PHITS comes packaged with the ability to simulate neutron transport in neutron supermirror guides. In order to introduce similar capabilities for the other codes, a number of different approaches have been taken. For example, Ref. \cite{Gallmeir2009} included this ability in a patch to MCNPX \cite{mcnp}, which was later ported and extended in a patch to MCNP6.2 \cite{mcnp6} in Ref. \cite{magan2020}. In a similar way, a patch including supermirror physics was also created for Geant4 \cite{dijulio2020}.
Alternatively, it has been suggested to use a neutron ray-tracing program, such as McStas \cite{mcstas1,mcstas2} or VITESS \cite{vitess1,vitess2}, to simulate the slow neutron transport in an instrument guide alone and transfer the unreflected neutrons to one of the above mentioned Monte-Carlo codes via specialized tools \cite{klinkby2013,knudsen2014,klinkby2014,kittelmann2017}.

The above mentioned approaches however imply a rather simplistic model of the interaction of the neutrons with the supermirror. Namely, a parameterized reflection probability is attributed to a single surface. The un-reflected neutrons are simply transferred to the supermirror layers, which are treated as a homogeneous material. Recently, a quantum-mechanical treatment of the neutron scattering in the layers was presented in \cite{koletov2019}. 
An interesting conclusion of this work was that the rigorous treatment provided a significantly different behavior of the neutron absorption in the supermirror layers, compared to, for example, the implementation in PHITS. This approach has also been included in McStas in the components derived from the neutron event logger tool \cite{knudsen2014,koletov2020,koletov2019b}, in order to be used in radiation shielding calculations. Lately, the patch to MCNP6.2 was further updated \cite{maganUCANS} and the model for neutron interaction with a supermirror was improved to make it possible to reproduce the theoretically calculated neutron absorption rates.

In light of the above mentioned theoretical and computational developments, there has so far not been a measurement, as far as the authors know, of the neutron absorption levels in supermirror coatings presented in previous literature. For this reason, we carried out such a study and present the results below. In the following sections, we give an overview of the experimental setup, followed by the data analysis methodology, and lastly present the results with comparison to the theoretical treatment presented in \cite{koletov2019}.

\section{Experimental Setup and Data Analysis}
\subsection{Experimental Setup}
The measurements were carried out using the SuperADAM instrument \cite{superADAM} at the Institut Laue-Langevin (ILL) \cite{ILL}, which provides a monochromatic beam of neutrons for reflectometry studies. Fig. 1 shows a schematic of the experimental setup. At the sample position, we placed a set of Ti and Ni/Mo $m=3$ and $m=4$ supermirrors and exposed them to a neutron beam of wavelength 5.2 Å. The samples were development test mirrors from Mirrotron Ltd. \cite{mirrotron}. The neutron beam illuminated the reflective side of both of these samples, where each was approximately 111 mm in length, 40 mm in width, and just under 4.0 mm thick. The Ni/Mo and Ti layers of the samples were deposited on a substrate of borofloat\textsuperscript{\textregistered} \cite{borofloat} glass. The samples were placed on a holder, comprising of boral and aluminium, at the sample position of the instrument and two slits placed upstream defined a beam of horizontal divergence of 0.2 mrad, resulting in a width of 0.6 mm at the sample position. Additionally, a 99.2 $\%$ pure Ti block, of dimensions of about 102 mm in length, 30 mm in width, and 2 mm thick, was used to help characterize the relative efficiency of the gamma-ray detector, as described below. The block was measured in the same geometry as the mirrors. Lastly, we exposed a borofloat\textsuperscript{\textregistered} substrate to the neutron beam, which provided the absolute normalization and a correction for several other effects, as described in detail below. Reflected neutrons were detected with a $^3$He detector placed downstream of the sample position, which was used to help ensure sample alignment. A detector upstream of the experiment was used to monitor the stability of the incoming beam.

Emitted gamma-rays from the samples were recorded with a high-purity germanium (HPGe) \cite{canberra} detector, where the center of the crystal was placed 10$\pm$2 cm from the reflective surface of the sample. The window of the HPGe detector was covered with a sheet of Li absorber to prevent any scattered neutrons from reaching the Ge volume and thereby creating background signals. The entire setup was enclosed within a customized lead shielding which ensured that any photons originating outside of the setup would have to cross approximately 10 cm of lead. This ensured the collection of low-background gamma-ray spectra. For each measurement, the emitted gamma-rays, along with the reflected neutrons, were recorded as a function of the angle of incidence of the neutron beam on the sample.

\subsection{Experimental Data and Analysis}
The data analysis was carried out by extracting gamma-ray peak counts and applying several corrections to the data. Each will be discussed next. Data on the energies and gamma-ray production from neutron capture were taken from \cite{Atlas} and the analysis was carried out using custom developed ROOT-based software \cite{root}.

A selection of the gamma-ray spectra used for the analysis are shown in Fig. \ref{Fig_Gammas}. These spectra are the sum over all the angles scanned for the $m=3$, $m=4$ supermirros and the substrate. As seen in the figure, a number of gamma-ray lines were observed. We were able to identify lines arising from the neutron capture in the Ni and Ti layers and for Al and Si, which can arise from capture in the sample holder or substrate. For the analysis, the relevant lines are indicated in the figure, which include the 478 keV line from capture in B, the 1382 keV line from capture in Ti, and the 8998 keV line from capture in Ni. These three lines were extracted as a function of the angle of incidence of the neutron beam. One can see that the 1382 keV and the 8998 keV lines do not appear in the substrate measurement, indicating they arise solely from the capture in the supermirror layers. The 478 keV line from B was used for the absolute normalization. 

After extracting the counts for each gamma-ray line, they were converted to the amount of neutron capture. The relevant data for this process is shown in Table 1. $\sigma_\gamma^{E_\gamma}$ is the partial capture cross-section for the indicated element and gamma-ray, while $\sigma_\gamma^{Tot}$ is the total capture cross-section for the indicated element. The number of gamma-rays per neutron capture is given by the quantity $P(E_\gamma)=\sigma_\gamma^{E_\gamma}/\sigma_\gamma^{Tot}$ and thus the counts can be converted to the amount of neutron capture through the relation $N_a=A_{E_\gamma}/P(E_\gamma)/\epsilon_{E_\gamma}$, where $N_a$ represents the neutron capture, $A_{E_\gamma}$ the peak counts and $\epsilon_{E_\gamma}$ is the detector efficiency. In this case, only the relative efficiency is needed, and the absolute normalization procedure was carried out as described in the following paragraphs.

Due to the large thermal neutron absorption cross-section of 764 barns for B, at the neutron wavelength used in this measurement, we calculated that around 3$\%$ of the neutrons would be lost to other processes besides capture in B, based on the compositions in \cite{boffy}. Thus one can consider that nearly the full beam is absorbed by B in the substrate. The fraction of neutrons absorbed per incident neutron in the Ni and Ti layers could then be calculated using $f^{Ni,Ti}_a=N_a^{Ni,Ti}/N_a^{B}$. The 3$\%$ systematic effect was additionally included in the error analysis.

The absolute normalization procedure was carried out by combining and extrapolating between two boron capture data sets, which are shown in Fig. \ref{Fig_Norm}, in addition to the constructed normalization curve and the $m=3$ data. In the figure, The relative count rates are plotted versus $m=q_z/q_c^{Ni}$, with $q_c^{Ni}$=0.0218 Å$^{-1}$ \cite{koletov2019}, and $q_z=(4\pi/\lambda)$sin$\theta$, where $\theta$ is the incidence angle and $\lambda$ is the wavelength of the incidence neutrons.
The two data sets used for the procedure include: 1) the data from the substrate up to just before the $m=4$ supermirror cutoff and 2) the data from the $m=4$ supermirror measurements above the cutoff. Above the cutoff, nearly all neutrons that are incident on the supermirror enter the substrate, and result in absorption by B. By using these datasets, it was possible to construct the absolute normalization factor across a larger range of incidence angles and at the same time normalizing out several combined effects. The first effect was due to the changing beam size footprint seen by the gamma-ray detector, as the setup was rotated. The second effect was that at higher rotational angles, above the $m=4$ cutoff, we observed a decrease in the gamma-ray intensities, which was associated with the neutron beam hitting the lead collimator around the sample. This was confirmed by the observation of the lead gamma ray at 7368 keV \cite{Atlas} above the $m$=4 cutoff and its intensity increasing simultaneously with the decrease of the intensities for gamma rays from the supermirror materials. The position of the lead line is indicated in Fig. 2 for reference. 

Two additional observations can be mentioned in reference to Fig. \ref{Fig_Norm}. The first is that we observed an approximately 5$\%$ greater maximum absorption in the B when the beam was on the substrate compared to when it was on the $m=3,4$ supermirrors above their respective cutoff values. A possible explanation could be related to scattering processes occurring in the supermirror layer before the neutrons reach the substrates. For example, Fig. 2 in \cite{koletov2019} suggests that above the supermirror cutoff, around 5\% of the events incident on the supermirror layers result in other processes, such as diffuse scattering, reflection and absorption instead of penetration through the layers. Thus we think one could consider an additional 5\% systematic effect in this region. The second is related to a possible additional constant contamination of the 478 keV line from elsewhere than the substrate. However, we observed the contribution from this line to be on the 1$\%$ level at the maximum measured reflectively of the supermirrors. 

The relative efficiencies used in the analysis were produced by using measured peak areas from both the Ti block and $m=4$ data sets, combined with the information in Ref. \cite{Atlas}. The relative efficiency of the 1382 keV Ti line, compared to the 478 keV B line, was taken from the extracted Ti data points. The efficiency of the 478 keV line was extrapolated from a fit to the low-energy Ti data points. The relative efficiency for the Ni line at 8998 keV, compared to the B line, was calculated by assuming the same relative efficiencies for the Ni line at 6838 keV and the Ti line at 6764 keV. A systematic error was included in the final result by varying this matching point within the up and lower error bars of those two points. Fig. \ref{Fig_Ref} shows the relative efficiencies used for the analysis, in addition to several other extracted lines from the measured data.

\section{Results and Discussion}
 Fig. \ref{Fig_M} shows the results from the measurements compared to recent calculations from \cite{koletov2019} for NiMo/Ti supermirrors.  Overall, it can be seen that the general behavior of the measurements is well reproduced by the calculations. A linear rising trend is observed with a sharp fall off once the supermirror critical angle has been reached. The calculated data for $m=3$ with Ti absorption and the $m=4$ with Ni absorption are in reasonably good agreement with the measurements, while the $m=4$ Ti absorption data is systematically higher in the measurements and we see the opposite effect for the $m=3$ Ni absorption data. In our analysis described above, there are indications that the samples we studied did not perform the same as those described in \cite{koletov2019}, which could be a part of the reason for the observed differences. For example, Fig. \ref{Fig_Norm} indicates an absorption of around 20\% for the m=3 supermirror and 30\% for the m=4 supermirror around the cutoff value. These are more than the values that would be anticipated from the reflectively data presented in \cite{koletov2019}. 
 
 Even considering the differences described above, the measured results do confirm the overall behavior of the neutron absorption, and the calculations deviate maximally about $50\%$ from the measurements. This level of agreement is typically suitable for Monte-Carlo shielding simulations and suggests that the approach presented in \cite{koletov2019} gives reasonable estimates of the neutron absorption in supermirror coatings. A more detailed study with a new set of supermirrors, where agreement between the performance of the calculated and measured supermirrors could be verified, would be of great benefit to help alleviate some of the challenges described here. Additionally, the calculations presented in \cite{koletov2019} covered $m$=2 to $m$=6 supermirrors. For $m=1$ mirrors, the waviness of the mirrors dominates the neutron losses, as described in \cite{koletov2019b,koletov2020}, and additional measurements on $m$=1 mirrors could be of interest to compare to the theoretical work described within the indicated references.  

\section{Conclusion}
In summary, we have reported on the first measurements of neutron absorption in supermirror coatings. The measurements were carried out on $m=3$ and $m=4$ supermirrors and the results were compared to recent theoretical calculations. Some of the challenges observed during the comparison were pointed out. Future work could be to carry out a new study where better agreement between the performance of the calculated and measured supermirrors could be verified. Additionally, it could be of interest to compare the measured and simulated gamma-ray spectra, arising from capture in the supermirror layers, for more realistic shielding geometries. Even considering these differences, the comparisons were quite satisfactory and provide strong evidence that the methods developed in \cite{koletov2019} can give accurate representations of the gamma-ray production along neutron guides to be used for radiation shielding calculations.

\section*{Acknowledgements}
The authors would like to thank Ken Andersen, Phil Bentley, Feri Mezei, and Luca Zanini for valuable discussions related to various aspects of this work. The authors would like to also thank Mirrotron Ltd. for providing the development test mirrors used in the measurements.

\bibliography{mybibfile}

\begin{thebibliography}{10}
\expandafter\ifx\csname url\endcsname\relax
  \def\url#1{\texttt{#1}}\fi
\expandafter\ifx\csname urlprefix\endcsname\relax\def\urlprefix{URL }\fi
\expandafter\ifx\csname href\endcsname\relax
  \def\href#1#2{#2} \def\path#1{#1}\fi

\bibitem{Garoby_2017}
R.~Garoby, A.~Vergara, H.~Danared, I.~Alonso, E.~Bargallo, B.~Cheymol,
  C.~Darve, M.~Eshraqi, H.~Hassanzadegan, A.~Jansson, I.~Kittelmann,
  Y.~Levinsen, M.~Lindroos, C.~Martins, {\O}.~Midttun, R.~Miyamoto, S.~Molloy,
  D.~Phan, A.~Ponton, E.~Sargsyan, T.~Shea, A.~Sunesson, L.~Tchelidze,
  C.~Thomas, M.~Jensen, W.~Hees, P.~Arnold, M.~Juni-Ferreira, F.~Jensen,
  A.~Lundmark, D.~McGinnis, N.~Gazis, J.~W. II, M.~Anthony, E.~Pitcher,
  L.~Coney, M.~Gohran, J.~Haines, R.~Linander, D.~Lyngh, U.~Oden, H.~Carling,
  R.~Andersson, S.~Birch, J.~Cereijo, T.~Friedrich, T.~Korhonen, E.~Laface,
  M.~Mansouri-Sharifabad, A.~Monera-Martinez, A.~Nordt, D.~Paulic, D.~Piso,
  S.~Regnell, M.~Zaera-Sanz, M.~Aberg, K.~Breimer, K.~Batkov, Y.~Lee,
  L.~Zanini, M.~Kickulies, Y.~Bessler, J.~Ringn{\'{e}}r, J.~Jurns,
  A.~Sadeghzadeh, P.~Nilsson, M.~Olsson, J.-E. Presteng, H.~Carlsson,
  A.~Polato, J.~Harborn, K.~Sjögreen, G.~Muhrer, F.~Sordo,
  \href{https://doi.org/10.1088/1402-4896/aa9bff}{The european spallation
  source design}, Physica Scripta 93~(1) (2017) 014001.
\newblock \href {http://dx.doi.org/10.1088/1402-4896/aa9bff}
  {\path{doi:10.1088/1402-4896/aa9bff}}.
\newline\urlprefix\url{https://doi.org/10.1088/1402-4896/aa9bff}

\bibitem{zanini2019}
L.~Zanini, K.~Andersen, K.~Batkov, E.~Klinkby, F.~Mezei, T.~Schönfeldt,
  A.~Takibayev,
  \href{https://www.sciencedirect.com/science/article/pii/S0168900219300087}{Design
  of the cold and thermal neutron moderators for the european spallation
  source}, Nuclear Instruments and Methods in Physics Research Section A:
  Accelerators, Spectrometers, Detectors and Associated Equipment 925 (2019)
  33--52.
\newblock \href {http://dx.doi.org/https://doi.org/10.1016/j.nima.2019.01.003}
  {\path{doi:https://doi.org/10.1016/j.nima.2019.01.003}}.
\newline\urlprefix\url{https://www.sciencedirect.com/science/article/pii/S0168900219300087}

\bibitem{Mezei1976}
F.~Mezei, Commun. Phys. (London), 1 (1976) 81.

\bibitem{Schanzer2016}
C.~Schanzer, M.~Schneider, P.~Böni,
  \href{https://doi.org/10.1088/1742-6596/746/1/012024}{Neutron optics: Towards
  applications for hot neutrons}, Journal of Physics: Conference Series 746
  (2016) 012024.
\newblock \href {http://dx.doi.org/10.1088/1742-6596/746/1/012024}
  {\path{doi:10.1088/1742-6596/746/1/012024}}.
\newline\urlprefix\url{https://doi.org/10.1088/1742-6596/746/1/012024}

\bibitem{cherkashyna2014}
N.~Cherkashyna, et~al., Overcoming high energy backgrounds at pulsed spallation
  sources, Proc. of the XXI Meeting of the International Collaboration on
  Advanced Neutron Sources (ICANS-XXI).

\bibitem{Santoro:2018fid}
V.~Santoro, D.~D. DiJulio, S.~Ansell, N.~Cherkashyna, G.~Muhrer, P.~M. Bentley,
  {Study of neutron shielding collimators for curved beamlines at the European
  Spallation Source}, J. Phys. Conf. Ser. 1046~(1) (2018) 012010.
\newblock \href {http://arxiv.org/abs/1804.02889} {\path{arXiv:1804.02889}},
  \href {http://dx.doi.org/10.1088/1742-6596/1046/1/012010}
  {\path{doi:10.1088/1742-6596/1046/1/012010}}.

\bibitem{background}
V.~Santoro, X.~X. Cai, D.~D. DiJulio, S.~Ansell, P.~M. Bentley, In-beam
  background suppression shield 18 (2015) 135--144.
\newblock \href {http://dx.doi.org/10.3233/JNR-160034}
  {\path{doi:10.3233/JNR-160034}}.

\bibitem{ersez2006}
T.~Ersez, G.~Braoudakis, J.~Osborn,
  \href{https://www.sciencedirect.com/science/article/pii/S0921452606013561}{Radiation
  shielding for neutron guides}, Physica B: Condensed Matter 385-386 (2006)
  1268--1270.
\newblock \href {http://dx.doi.org/https://doi.org/10.1016/j.physb.2006.06.028}
  {\path{doi:https://doi.org/10.1016/j.physb.2006.06.028}}.
\newline\urlprefix\url{https://www.sciencedirect.com/science/article/pii/S0921452606013561}

\bibitem{jensen2014}
C.~Cooper-Jensen, et~al., "m=1" coatings for neutron guides, J. Phys.: Conf.
  Ser. 528 (2014) 012005.

\bibitem{niita2006}
K.~Niita, T.~Sato, H.~Iwase, H.~Nose, H.~Nakashima, L.~Sihver,
  \href{https://www.sciencedirect.com/science/article/pii/S1350448706001351}{Phits—a
  particle and heavy ion transport code system}, Radiation Measurements 41~(9)
  (2006) 1080--1090, space Radiation Transport, Shielding, and Risk Assessment
  Models.
\newblock \href
  {http://dx.doi.org/https://doi.org/10.1016/j.radmeas.2006.07.013}
  {\path{doi:https://doi.org/10.1016/j.radmeas.2006.07.013}}.
\newline\urlprefix\url{https://www.sciencedirect.com/science/article/pii/S1350448706001351}

\bibitem{mcnp}
\href{https://mcnp.lanl.gov/}{Los alamos national laboratory, mcnp home page}.
\newline\urlprefix\url{https://mcnp.lanl.gov/}

\bibitem{Geant4}
\href{http://www.geant4.org}{Geant4}.
\newline\urlprefix\url{http://www.geant4.org}

\bibitem{agostinelli2003}
S.~Agostinelli, J.~Allison, K.~Amako, J.~Apostolakis, H.~Araujo, P.~Arce,
  M.~Asai, D.~Axen, S.~Banerjee, G.~Barrand, F.~Behner, L.~Bellagamba,
  J.~Boudreau, L.~Broglia, A.~Brunengo, H.~Burkhardt, S.~Chauvie, J.~Chuma,
  R.~Chytracek, G.~Cooperman, G.~Cosmo, P.~Degtyarenko, A.~Dell'Acqua,
  G.~Depaola, D.~Dietrich, R.~Enami, A.~Feliciello, C.~Ferguson, H.~Fesefeldt,
  G.~Folger, F.~Foppiano, A.~Forti, S.~Garelli, S.~Giani, R.~Giannitrapani,
  D.~Gibin, J.~{Gómez Cadenas}, I.~González, G.~{Gracia Abril}, G.~Greeniaus,
  W.~Greiner, V.~Grichine, A.~Grossheim, S.~Guatelli, P.~Gumplinger,
  R.~Hamatsu, K.~Hashimoto, H.~Hasui, A.~Heikkinen, A.~Howard, V.~Ivanchenko,
  A.~Johnson, F.~Jones, J.~Kallenbach, N.~Kanaya, M.~Kawabata, Y.~Kawabata,
  M.~Kawaguti, S.~Kelner, P.~Kent, A.~Kimura, T.~Kodama, R.~Kokoulin,
  M.~Kossov, H.~Kurashige, E.~Lamanna, T.~Lampén, V.~Lara, V.~Lefebure,
  F.~Lei, M.~Liendl, W.~Lockman, F.~Longo, S.~Magni, M.~Maire, E.~Medernach,
  K.~Minamimoto, P.~{Mora de Freitas}, Y.~Morita, K.~Murakami, M.~Nagamatu,
  R.~Nartallo, P.~Nieminen, T.~Nishimura, K.~Ohtsubo, M.~Okamura, S.~O'Neale,
  Y.~Oohata, K.~Paech, J.~Perl, A.~Pfeiffer, M.~Pia, F.~Ranjard, A.~Rybin,
  S.~Sadilov, E.~{Di Salvo}, G.~Santin, T.~Sasaki, N.~Savvas, Y.~Sawada,
  S.~Scherer, S.~Sei, V.~Sirotenko, D.~Smith, N.~Starkov, H.~Stoecker,
  J.~Sulkimo, M.~Takahata, S.~Tanaka, E.~Tcherniaev, E.~{Safai Tehrani},
  M.~Tropeano, P.~Truscott, H.~Uno, L.~Urban, P.~Urban, M.~Verderi, A.~Walkden,
  W.~Wander, H.~Weber, J.~Wellisch, T.~Wenaus, D.~Williams, D.~Wright,
  T.~Yamada, H.~Yoshida, D.~Zschiesche,
  \href{https://www.sciencedirect.com/science/article/pii/S0168900203013688}{Geant4—a
  simulation toolkit}, Nuclear Instruments and Methods in Physics Research
  Section A: Accelerators, Spectrometers, Detectors and Associated Equipment
  506~(3) (2003) 250--303.
\newblock \href
  {http://dx.doi.org/https://doi.org/10.1016/S0168-9002(03)01368-8}
  {\path{doi:https://doi.org/10.1016/S0168-9002(03)01368-8}}.
\newline\urlprefix\url{https://www.sciencedirect.com/science/article/pii/S0168900203013688}

\bibitem{fluka}
\href{http://www.fluka.org/}{The official fluka site: Fluka home}.
\newline\urlprefix\url{http://www.fluka.org/}

\bibitem{Gallmeir2009}
F.~X. Gallmeier, M.~Wohlmuther, U.~Filges, D.~Kiselev, G.~Muhrer,
  \href{https://doi.org/10.13182/NT09-A9304}{Implementation of neutron mirror
  modeling capability into mcnpx and its demonstration in first applications},
  Nuclear Technology 168~(3) (2009) 768--772.
\newblock \href {http://arxiv.org/abs/https://doi.org/10.13182/NT09-A9304}
  {\path{arXiv:https://doi.org/10.13182/NT09-A9304}}, \href
  {http://dx.doi.org/10.13182/NT09-A9304} {\path{doi:10.13182/NT09-A9304}}.
\newline\urlprefix\url{https://doi.org/10.13182/NT09-A9304}

\bibitem{mcnp6}
T.~Goorley, M.~James, T.~Booth, F.~Brown, J.~Bull, L.~J. Cox, J.~Durkee,
  J.~Elson, M.~Fensin, R.~A. Forster, J.~Hendricks, H.~G. Hughes, R.~Johns,
  B.~Kiedrowski, R.~Martz, S.~Mashnik, G.~McKinney, D.~Pelowitz, R.~Prael,
  J.~Sweezy, L.~Waters, T.~Wilcox, T.~Zukaitis,
  \href{https://doi.org/10.13182/NT11-135}{Initial mcnp6 release overview},
  Nuclear Technology 180~(3) (2012) 298--315.
\newblock \href {http://arxiv.org/abs/https://doi.org/10.13182/NT11-135}
  {\path{arXiv:https://doi.org/10.13182/NT11-135}}, \href
  {http://dx.doi.org/10.13182/NT11-135} {\path{doi:10.13182/NT11-135}}.
\newline\urlprefix\url{https://doi.org/10.13182/NT11-135}

\bibitem{magan2020}
M.~Magán, R.~M. Bergmann,
  \href{https://www.sciencedirect.com/science/article/pii/S0168900219314755}{Supermirror
  physics with event biasing in mcnp6}, Nuclear Instruments and Methods in
  Physics Research Section A: Accelerators, Spectrometers, Detectors and
  Associated Equipment 955 (2020) 163168.
\newblock \href {http://dx.doi.org/https://doi.org/10.1016/j.nima.2019.163168}
  {\path{doi:https://doi.org/10.1016/j.nima.2019.163168}}.
\newline\urlprefix\url{https://www.sciencedirect.com/science/article/pii/S0168900219314755}

\bibitem{dijulio2020}
D.~DiJulio, et~al., Simulating neutron transport in long beamlines at a
  spallation neutron source using geant4, J. Neutron Res. 1 (2020) 183.

\bibitem{mcstas1}
K.~Lefmann, K.~Nielsen,
  \href{https://doi.org/10.1080/10448639908233684}{Mcstas, a general software
  package for neutron ray-tracing simulations}, Neutron News 10~(3) (1999)
  20--23.
\newblock \href
  {http://arxiv.org/abs/https://doi.org/10.1080/10448639908233684}
  {\path{arXiv:https://doi.org/10.1080/10448639908233684}}, \href
  {http://dx.doi.org/10.1080/10448639908233684}
  {\path{doi:10.1080/10448639908233684}}.
\newline\urlprefix\url{https://doi.org/10.1080/10448639908233684}

\bibitem{mcstas2}
P.~Willendrup, et~al., J. Phys.:Conf. Ser. 528 (2014) 012035.

\bibitem{vitess1}
K.~Lieutenant, G.~Zsigmond, S.~Manoshin, M.~Fromme, H.~N. Bordallo,
  D.~Champion, J.~Peters, F.~Mezei,
  \href{https://doi.org/10.1117/12.562814}{{Neutron instrument simulation and
  optimization using the software package VITESS}}, in: M.~S. del Rio (Ed.),
  Advances in Computational Methods for X-Ray and Neutron Optics, Vol. 5536,
  International Society for Optics and Photonics, SPIE, 2004, pp. 134 -- 145.
\newblock \href {http://dx.doi.org/10.1117/12.562814}
  {\path{doi:10.1117/12.562814}}.
\newline\urlprefix\url{https://doi.org/10.1117/12.562814}

\bibitem{vitess2}
C.~Zendler, K.~Lieutenant, D.~Nekrassov, M.~Fromme,
  \href{https://doi.org/10.1088/1742-6596/528/1/012036}{{VITESS} 3
  {\textendash} virtual instrumentation tool for the european spallation
  source}, Journal of Physics: Conference Series 528 (2014) 012036.
\newblock \href {http://dx.doi.org/10.1088/1742-6596/528/1/012036}
  {\path{doi:10.1088/1742-6596/528/1/012036}}.
\newline\urlprefix\url{https://doi.org/10.1088/1742-6596/528/1/012036}

\bibitem{klinkby2013}
E.~Klinkby, B.~Lauritzen, E.~Nonbøl, P.~{Kjær Willendrup}, U.~Filges,
  M.~Wohlmuther, F.~X. Gallmeier,
  \href{https://www.sciencedirect.com/science/article/pii/S0168900212011874}{Interfacing
  mcnpx and mcstas for simulation of neutron transport}, Nuclear Instruments
  and Methods in Physics Research Section A: Accelerators, Spectrometers,
  Detectors and Associated Equipment 700 (2013) 106--110.
\newblock \href {http://dx.doi.org/https://doi.org/10.1016/j.nima.2012.10.052}
  {\path{doi:https://doi.org/10.1016/j.nima.2012.10.052}}.
\newline\urlprefix\url{https://www.sciencedirect.com/science/article/pii/S0168900212011874}

\bibitem{knudsen2014}
E.~{Bergbäck Knudsen}, E.~{Bryndt Klinkby}, P.~{Kjær Willendrup},
  \href{https://www.sciencedirect.com/science/article/pii/S0168900213016331}{Mcstas
  event logger: Definition and applications}, Nuclear Instruments and Methods
  in Physics Research Section A: Accelerators, Spectrometers, Detectors and
  Associated Equipment 738 (2014) 20--24.
\newblock \href {http://dx.doi.org/https://doi.org/10.1016/j.nima.2013.11.071}
  {\path{doi:https://doi.org/10.1016/j.nima.2013.11.071}}.
\newline\urlprefix\url{https://www.sciencedirect.com/science/article/pii/S0168900213016331}

\bibitem{klinkby2014}
E.~Klinkby, et~al., J. Phys.:Conf. Ser. 528 (2014) 012032.

\bibitem{kittelmann2017}
T.~Kittelmann, E.~Klinkby, E.~Knudsen, P.~Willendrup, X.~Cai, K.~Kanaki,
  \href{https://www.sciencedirect.com/science/article/pii/S0010465517301261}{Monte
  carlo particle lists: Mcpl}, Computer Physics Communications 218 (2017)
  17--42.
\newblock \href {http://dx.doi.org/https://doi.org/10.1016/j.cpc.2017.04.012}
  {\path{doi:https://doi.org/10.1016/j.cpc.2017.04.012}}.
\newline\urlprefix\url{https://www.sciencedirect.com/science/article/pii/S0010465517301261}

\bibitem{koletov2019}
R.~Kolevatov, C.~Schanzer, P.~Böni,
  \href{https://www.sciencedirect.com/science/article/pii/S0168900218318837}{Neutron
  absorption in supermirror coatings: Effects on shielding}, Nuclear
  Instruments and Methods in Physics Research Section A: Accelerators,
  Spectrometers, Detectors and Associated Equipment 922 (2019) 98--107.
\newblock \href {http://dx.doi.org/https://doi.org/10.1016/j.nima.2018.12.069}
  {\path{doi:https://doi.org/10.1016/j.nima.2018.12.069}}.
\newline\urlprefix\url{https://www.sciencedirect.com/science/article/pii/S0168900218318837}

\bibitem{koletov2020}
R.~Kolevatov, J. Synch. Investig. 14 (2020) S105--S107.

\bibitem{koletov2019b}
R.~Koletov, J. Neut. Res. 21 (2019) 79--85.

\bibitem{maganUCANS}
{M. Magán, R. M. Bergmann, and O. González}, {Detailed Supermirror physics in
  MCNP6, 23rd meeting of the International Collaboration on Advanced Neutron
  Sources (ICANS XXIII), Chattanooga, Tennessee} (2019).

\bibitem{superADAM}
A.~Devishvili, K.~Zhernenkov, A.~J.~C. Dennison, B.~P. Toperverg, M.~Wolff,
  B.~Hjörvarsson, H.~Zabel,
  \href{https://doi.org/10.1063/1.4790717}{Superadam: Upgraded polarized
  neutron reflectometer at the institut laue-langevin}, Review of Scientific
  Instruments 84~(2) (2013) 025112.
\newblock \href {http://arxiv.org/abs/https://doi.org/10.1063/1.4790717}
  {\path{arXiv:https://doi.org/10.1063/1.4790717}}, \href
  {http://dx.doi.org/10.1063/1.4790717} {\path{doi:10.1063/1.4790717}}.
\newline\urlprefix\url{https://doi.org/10.1063/1.4790717}

\bibitem{ILL}
\href{https://www.ill.eu/}{Ill neutrons for society}.
\newline\urlprefix\url{https://www.ill.eu/}

\bibitem{mirrotron}
\href{https://mirrotron.com/}{Mirrotron}.
\newline\urlprefix\url{https://mirrotron.com/}

\bibitem{borofloat}
\href{https://www.schott.com/en-us/products/borofloat}{Borofloat\textsuperscript{\textregistered}
  {SCHOTT}}.
\newline\urlprefix\url{https://www.schott.com/en-us/products/borofloat}

\bibitem{canberra}
\href{https://www.mirion.com/}{Mirion technologies, inc. radiation measurement
  and detection devices}.
\newline\urlprefix\url{https://www.mirion.com/}

\bibitem{Atlas}
Database of Prompt Gamma Rays From Slow Neutron Capture for Elemental Analysis,
  Report number: STI/PUB/1263, International Atomic Energy Agency, Vienna,
  2007.

\bibitem{root}
\href{https://root.cern.ch}{Root: analyzing petabytes of data, scientifically.}
\newline\urlprefix\url{https://root.cern.ch}

\bibitem{boffy}
R.~Boffy, {DESIGN OF A NEW NEUTRON DELIVERY SYSTEM FOR HIGH FLUX SOURCE, PhD
  Thesis, Polytechnic University of Madrid} (2016).

\end{thebibliography}
\newpage

\begin{table}[h]
  \caption{Data used for the analysis of the gamma-ray counts, taken from \cite{Atlas}. $\sigma_\gamma^{E_\gamma}$ is the partial capture cross-section for the indicated gamma-ray line and element and $\sigma_\gamma^{Tot}$ is total capture cross-section for the indicated element. }
\begin{center}
  \small
  \begin{tabular}{|l|l|l|l|}
    \hline
    Element & Gamma-ray Energy & $\sigma_\gamma^{E_\gamma}$ &  $\sigma_\gamma^{Tot}$ \\
    \hline
   B  & 478 keV  & 716(25) b  & 764(25) b \\
   Ti & 1382 keV & 5.18(12) b & 6.08(19) b \\
   Ni & 8998 keV & 1.49(3) b  & 4.39(15) b\\
\hline
\end{tabular}
\end{center}
\label{default}
\end{table}%

\begin{figure}[t]
\resizebox{1.2\textwidth}{!}{\includegraphics{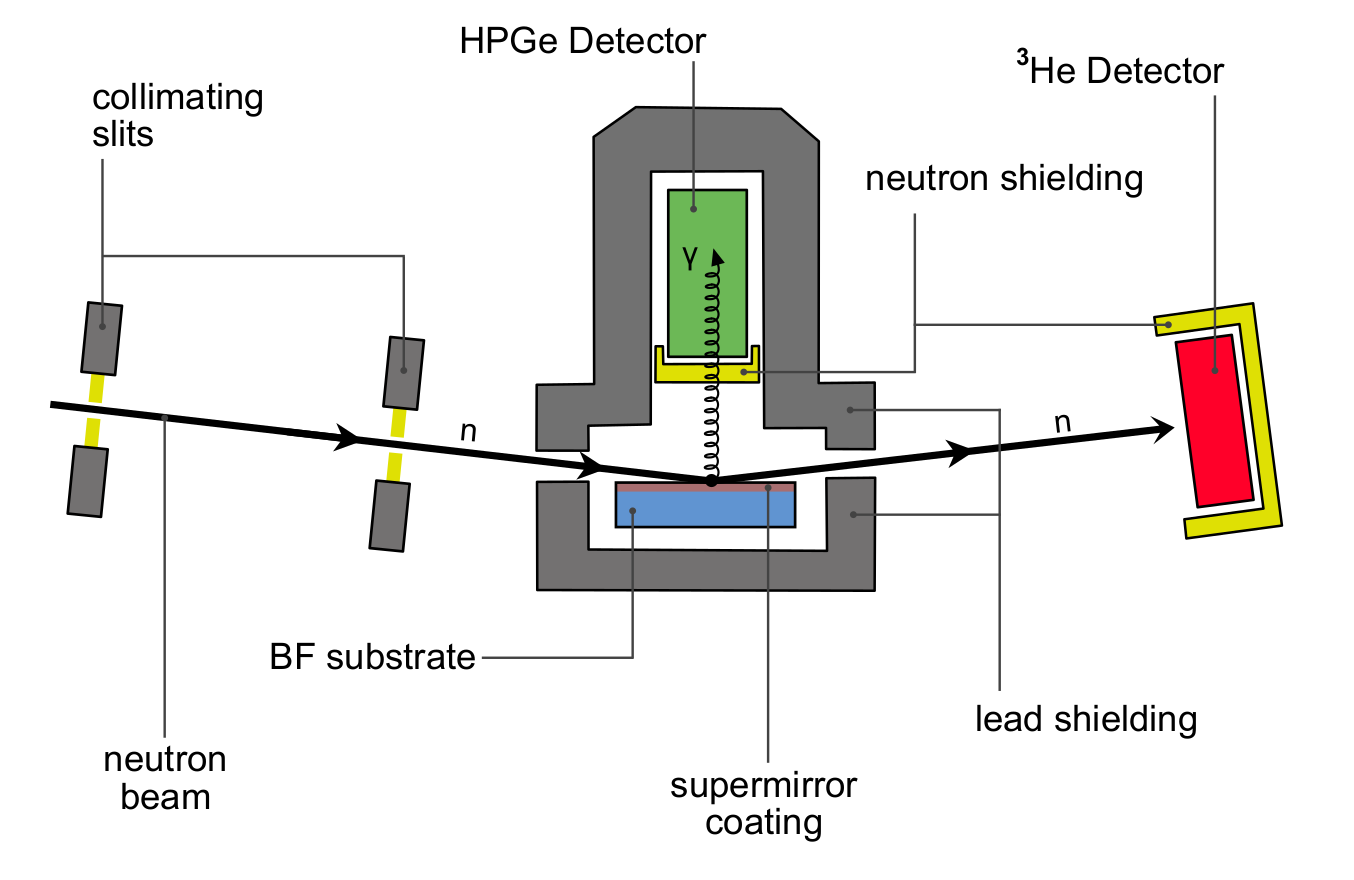}}
\caption{Schematic overview of the experimental setup.}\label{Fig_Exp}
\end{figure}
\begin{figure}[t]
\resizebox{1.2\textwidth}{!}{\includegraphics{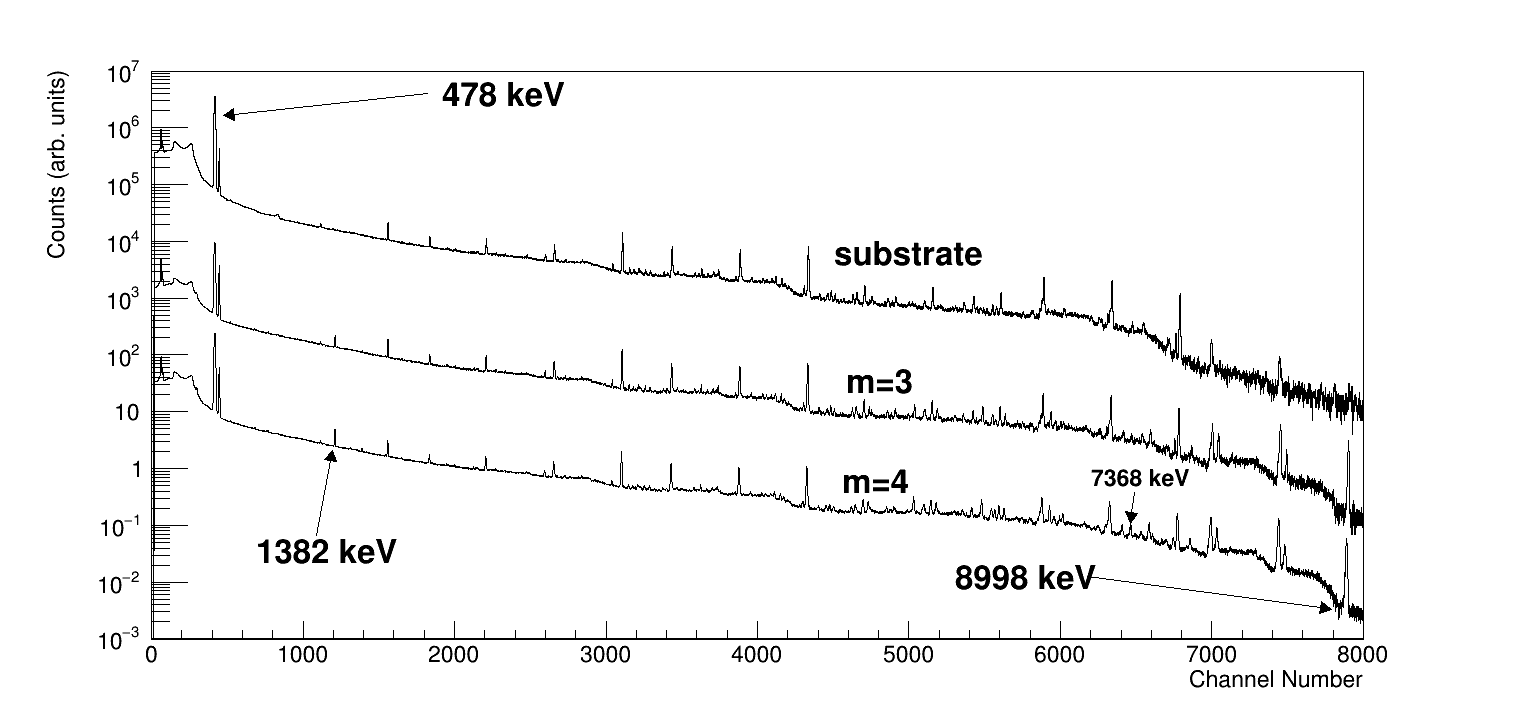}}
\caption{Selection of the gamma-ray spectra collected through the measurements. The relevant lines for the analysis are indicated. The 478 keV line arises from capture in boron in the substrate, the 1382 keV line arises from capture in Ti in the supermirror, the 8998 keV line arises from capture in Ni in the supermirror, and the 7368 keV line arises from capture in the lead collimator, as described in the text.}\label{Fig_Gammas}
\end{figure}

\begin{figure}[t]
\resizebox{1.2\textwidth}{!}{\includegraphics{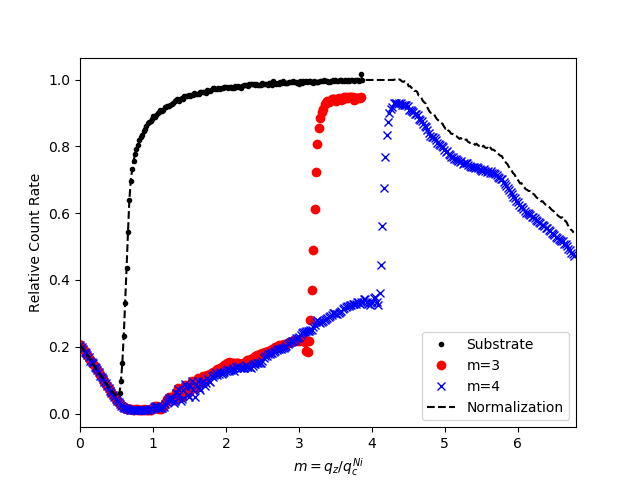}}
\caption{Relative count rates for the 478 keV boron capture line for the m=3 supermirror, m=4 supermirror, and the substrate. The constructed curve for normalization is also shown in the figure.}\label{Fig_Norm}
\end{figure}

\begin{figure}[t]
\resizebox{1.2\textwidth}{!}{\includegraphics{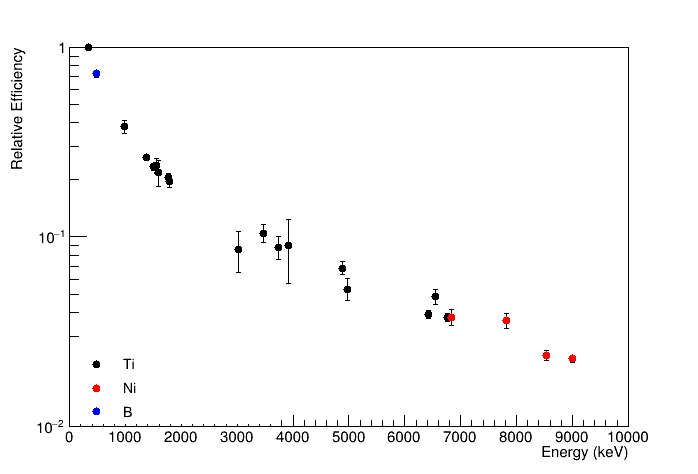}}
\caption{Relative gamma-ray efficiencies for the setup as calculated and described in the text. The Ni and Ti data comes from the measurements and the B data comes from an extrapolation between the low-energy measured data points.}\label{Fig_Ref}
\end{figure}

\begin{figure}[t]
\resizebox{1.2\textwidth}{!}{\includegraphics{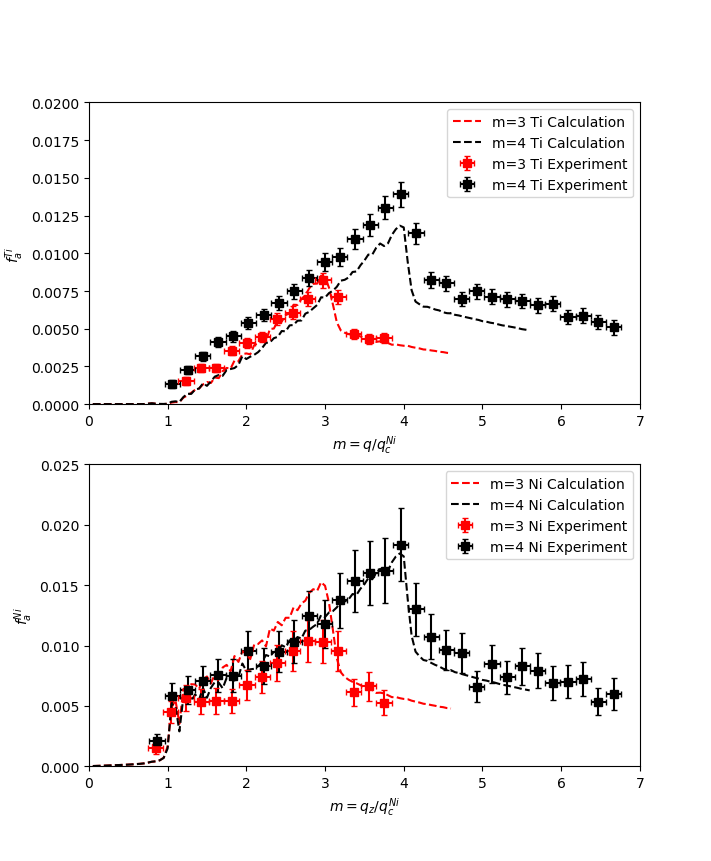}}
\caption{The fraction of neutrons absorbed in the supermirror layers for the measurements compared to the calculations from \cite{koletov2019}. The vertical error bars represent the estimates of the standard errors while the horizontal error bars represent the angular bin widths.}\label{Fig_M}
\end{figure}

\end{document}